\documentclass[11pt]{article}

\usepackage[T1]{fontenc}
\usepackage[utf8]{inputenc}
\usepackage{lmodern}
\usepackage{microtype}
\usepackage[letterpaper,margin=1in]{geometry}
\usepackage{amsmath}
\usepackage{graphicx}
\usepackage{booktabs}
\usepackage{array}
\usepackage{tabularx}
\usepackage{xcolor}
\usepackage{tikz}
\usetikzlibrary{shapes.geometric,shapes.misc,arrows.meta,calc,positioning}
\usepackage[hidelinks]{hyperref}

\newcommand{\zenodoDOI}{10.5281/zenodo.20531284}

\newif\ifshortintro
\shortintrotrue

\title{Operational Assurance for LLM-Based Systems\\under Governance Opacity}

\author{Pedro Pinacho-Davidson\\[2pt]
  \small Department of Informatics Engineering and Computer Science (DIICC)\\
  \small AI \& Cybersecurity Research Group\\
  \small Universidad de Concepción, Concepción, Chile\\
  \small \texttt{ppinacho@udec.cl}\quad ORCID: \href{https://orcid.org/0000-0001-9324-284X}{0000-0001-9324-284X}}

\date{}

\begin{document}
\maketitle

\begin{abstract}
In the contemporary artificial intelligence ecosystem, foundation models are increasingly deployed under AI-as-a-Service (AIaaS) schemes characterized by high technical opacity, technological concentration, and limited external auditing capability over governance, training, and internal operational processes. Under these conditions, traditional ethical oversight approaches that focus exclusively on corporate governance or on internal model inspection become insufficient to ensure effective control over AI systems deployed in real-world contexts.

This work proposes a pragmatic operational assurance framework inspired by Zero-Trust principles from cybersecurity, which shifts effective auditing efforts toward the application layer, where risk acquires observable, contextualized, and operationally verifiable properties. The proposal articulates two complementary evaluation levels: a black-box evaluation aimed at characterizing general behaviors, vulnerabilities, and risky capabilities of foundation models through conversational scanning, and a grey-box evaluation focused on the operational contextualization of risk within the actual deployment environment of the audited application.

The framework incorporates mechanisms to move from general, CVSS-inspired severity metrics toward contextualized operational risk assessments that take into account user profiles, industrial domains, and organizational knowledge. In addition, the proposal strengthens practical auditability through digitally signed explainability reports (signed XAI reports), designed to support verifiable traceability, continuous monitoring, and operational accountability in LLM-based sociotechnical systems.
\end{abstract}


\vspace{0.5em}
\noindent\fbox{\parbox{\dimexpr\linewidth-2\fboxsep-2\fboxrule}{\small
\textbf{Note.} This is the English version of the preprint originally published in Spanish as ``Aseguramiento Operacional para Sistemas Basados en LLM bajo Opacidad de Gobernanza'', Zenodo, 2026, \href{https://doi.org/\zenodoDOI}{doi:\zenodoDOI}. The content corresponds to the original, with minor editorial revisions.}}

\section{Context}
\label{sec:context}

\ifshortintro
Trust in global governance mechanisms over critical processes has eroded significantly, and multilateral institutions have shown clear limitations in coordinating responses to complex global challenges~\cite{weiss2014,zurn2018,guterres2026}. In an increasingly belligerent geopolitical environment, transparency in the governance and auditing of strategic technologies such as advanced AI models remains limited or opaque to international oversight~\cite{brundage2018,kroll2017}.
\fi

At the same time, technological development in AI is highly concentrated in a small number of large technology companies, which lead, almost exclusively, the research and deployment of foundation models. These systems are trained at massive scale on broad, unlabeled datasets, which endows them with emergent capabilities and allows them to adapt to a wide variety of downstream tasks~\cite{bommasani2021}. However, this competitive dynamic encourages opacity regarding architectures, training data, and the emergent capabilities of the models themselves~\cite{bommasani2021,crawford2021}. Likewise, the consolidation of the AI-as-a-Service (AIaaS) paradigm, in which models are offered as proprietary services, reinforces centralized control. Given the extremely high cost of developing these systems, their provision as a service creates technical and commercial dependencies at a global scale, inhibiting third-party development and straining ethical principles such as justice and equity by concentrating in a few actors the capacity to influence sociotechnical systems of global reach~\cite{bostrom2014,zuboff2019}.

This is the current context. Although future technical advances, particularly those aimed at democratizing AI development, may plausibly change this structure, in the short term it is necessary to adopt a pragmatic perspective focused on the safe application of these technologies as they are currently offered and exploited. From an ethical standpoint, proposals such as those of Luciano Floridi and colleagues call for auditing AI systems at three layers: (i) the governance of the organizations that develop foundation models, (ii) the technological product itself, for example language or multimodal models, and (iii) the application level, that is, the concrete uses of such systems in real contexts~\cite{floridi2018,floridi2019,mokander2023}.

However, the governance layer is dominated by what may be called ``soft ethics'': a set of non-binding principles that are rarely translated into effective regulation. The influence of industry on regulatory processes makes it possible, in many cases, to delay, adapt, or dilute regulatory frameworks in favor of established actors~\cite{cohen2019,eubanks2018}. In the current geopolitical scenario, it is difficult to foresee coordinated efforts capable of establishing auditable governance mechanisms that guarantee robust ethical standards.

The open-source ecosystem does not offer a clear order either. Platforms such as Hugging Face facilitate model distribution but exercise no effective control over model capabilities or their downstream uses. Although there are technical proposals aimed at improving model traceability, for example through cryptographic hashing or PKI-based infrastructures, these have not yet been consolidated as standards in the ecosystem. Consequently, beyond the governance challenges, the technological product itself may exhibit questionable levels of security and reliability~\cite{carlini2021,nist2023}.

\section{Distilling the Context}
\label{sec:distilling}

Under the current conditions of the AI ecosystem, it is reasonable to adopt perspectives inspired by Zero-Trust approaches from cybersecurity~\cite{rose2020}. In these models, no component is assumed to be inherently trustworthy; instead, every interaction must be continuously verified within the scopes where there is real capacity for oversight and control. Transferred to the AI context, this approach suggests that the greatest evaluation and auditing efforts should be concentrated ``downstream'', that is, at the level of the concrete implementation of the AI application. It is precisely at these levels that impacts cease to be hypothetical and materialize on real users, organizations, and processes. Unlike the upper layers, such as corporate governance or foundation models, the application level is embedded in operational regulatory frameworks, defined jurisdictions, social norms, and institutional mechanisms capable of responding to concrete harms.

From this perspective, risk acquires different properties depending on the layer under observation. At the governance and model levels, risks tend to be formulated abstractly and generally, associated with potential scenarios, emergent capabilities, or interactions that are difficult to anticipate. At the application layers, in contrast, adverse effects become observable, contextualized, and in many cases measurable. This enables approaches closer to traditional practices of risk management and technology oversight.

This view is also consistent with recent regulatory approaches such as the European Union AI Act, which establishes differentiated oversight mechanisms according to the risk level of the concrete use of the technology~\cite{euaiact2024}. Under this logic, not all systems require the same degree of control; rather, the depth of evaluation should correspond to the potential impact of each specific application. In this context, the security of AI systems cannot be understood solely as a technical problem, but as an operational manifestation of fundamental ethical principles~\cite{floridi2018,floridi2019,nist2023}. Beneficence relates to the legitimate and responsible intent of the services offered; non-maleficence to the mitigation of harms and side effects; autonomy to the preservation of human decision-making capacity; justice to the equitable distribution of impacts and capabilities; and explicability to the possibility of understanding, auditing, and challenging the behavior of the systems.

In particular, non-maleficence acquires a dimension closely tied to cybersecurity, since many of the practical risks associated with AI systems arise precisely from vulnerabilities, unforeseen malicious uses, adversarial manipulation, information leakage, or unexpected emergent behaviors~\cite{nist2023,goodfellow2018}. Similarly, explicability plays a cross-cutting role, as it enables the practical verification of the other ethical principles within real operating contexts~\cite{floridi2019}.

Although a structured three-level ethical evaluation of AI systems (governance, model, and application) has been proposed~\cite{floridi2018,floridi2019,mokander2023}, the contemporary reality of the technology ecosystem makes it difficult to implement robust and auditable oversight mechanisms spanning all of these levels. Under these conditions, it becomes necessary to reconsider where real evaluation, monitoring, and control capabilities can be exercised effectively. This is especially true for actors who consume AIaaS to build their own systems and who cannot demand the necessary level of scrutiny from provider companies, their processes, and their AI models.

\section{A Layered Reality}
\label{sec:layers}

In the context of a still immature technology such as LLMs, whose technical capabilities and fields of application continue to expand rapidly, cybersecurity cannot be understood as centered solely on the technical protection of digital infrastructures. From the perspective of this work, cybersecurity is an operational manifestation of applied ethics in information systems, aimed at mitigating the problems arising from the design, deployment, and use of technologies that do not adequately comply with fundamental ethical principles such as beneficence, non-maleficence, justice, autonomy, and explicability~\cite{floridi2018,floridi2019,nist2023}.

Under this perspective, the ethical auditability of AI systems has profoundly different properties and levels of feasibility depending on the level observed: governance, foundation model, or concrete application.

\subsection{Governance: The Organization and Its Processes}

The governance layer is centered mainly on organizational processes, internal policies, and strategic guidelines associated with the development of AI systems. At this level, beneficence can be evaluated by analyzing institutional missions, stated objectives, compliance mechanisms, and corporate accountability structures.

The autonomy of the people exposed to the AI system is harder to characterize at this layer, since it depends heavily on the final application contexts. However, policies related to the responsible design of human--machine interaction could be evaluated, particularly with respect to mechanisms that encourage excessive anthropomorphization of, or psychological dependence on, conversational systems. Non-maleficence can only be analyzed indirectly, mainly through guidelines for safe model development, internal evaluation protocols, and processes for managing foreseeable risks. In this context, potential harms remain speculative, because the final usage scenarios cannot yet be fully known.

Justice can be addressed through methodological inspection of data collection processes, dataset curation, and bias mitigation policies. However, as long as the final applications remain undetermined, it is difficult to assess the real impact of such measures on concrete groups of individuals. Finally, explicability at this layer relates mainly to organizational documentation and the traceability of development processes. Rather than directly explaining the behavior of the final system, documentation becomes relevant as a mechanism of institutional accountability and as support for subsequent audits at lower layers.

It is important to note that there are currently few regulatory mechanisms capable of imposing deep and verifiable audits at this level. In practice, self-regulation schemes grounded in ``soft ethics'' principles predominate, which introduces a significant degree of opacity regarding the depth, quality, and truthfulness of the review processes carried out by the technology providers themselves~\cite{cohen2019,hagendorff2020,mittelstadt2019}.

\subsection{The Foundation Model: The Technological Product}

At the foundation model layer, the object of evaluation is the technological product itself, independently of the specific applications that specialize or restrict its behavior.

Beneficence is particularly difficult to audit at this level, since foundation models are usually inherently multipurpose and agnostic with respect to their final uses. Consequently, determining whether the system generates concrete benefits depends largely on the subsequent implementation context. Non-maleficence is probably the main focus of evaluation at this layer. Here, general tests predominate, aimed at identifying vulnerabilities, emergent behaviors, adversarial manipulation, harmful content generation, or security failures. This type of evaluation usually relies on benchmarks, red teaming, and experimental characterization of capabilities, extending approaches similar to those found in model cards or system cards~\cite{bommasani2021,nist2023}.

Autonomy can be explored by analyzing interaction patterns that foster emotional dependence, anthropomorphization, or behavioral manipulation. Different user profiles can also be simulated to identify particular patterns of psychological manipulation to which users might be exposed. Justice is usually evaluated through bias analysis and differential behavior in hypothetical scenarios. However, in the absence of a concrete operational domain, these evaluations remain partially speculative and depend heavily on the scenarios selected for testing.

Explicability has significant limitations at this layer. Without higher-level orchestration mechanisms, such as advanced prompting, contextual retrieval, or external task-decomposition structures, the observable output is essentially the emergent behavior of a complex neural network. Moreover, since the final purposes of use are unknown, it is also uncertain which explanatory structures would actually be necessary or sufficient.

\subsection{The Application: Operational Impact}

The application layer is the level at which AI systems actually interact with users, organizations, and real processes. Consequently, it is also the level at which ethical principles become most observable in practice. Beneficence can be evaluated much more directly, since the concrete purpose of the system has already been defined and deployed in a specific operational context, where the foundation model becomes part of a broader sociotechnical system.

Non-maleficence becomes especially relevant because potential harms cease to be hypothetical and become measurable, contextualized risks. Since the operating environment is known, it becomes possible to identify specific side effects, attack surfaces, and concrete operational consequences. Autonomy extends the problems identified at the model level, but now considering real users and specific profiles of psychological susceptibility, behavioral manipulation, or functional dependence.

Justice can be evaluated directly on affected groups and concrete impact scenarios. In this context, different notions of algorithmic fairness can be applied, calibrating metrics and evaluating trade-offs according to the actual requirements of the application domain. Finally, explicability reaches its highest level of auditability at this layer. Unlike the isolated foundation model, applications incorporate additional layers of software, advanced prompting, information retrieval mechanisms, and logical control structures that can be inspected and monitored. It also becomes possible to implement traceability mechanisms comparable to non-repudiation systems, including verifiable logs, decision audits, and different levels of explanation tailored to both end users and technical auditors.

Table~\ref{tab:auditability} summarizes how ethical principles acquire different levels of observability, verifiability, and operationalization depending on the layer analyzed.

\begin{table}[t]
\centering
\caption{Ethical auditability characteristics by evaluation level.}
\label{tab:auditability}
\small
\renewcommand{\arraystretch}{1.25}
\begin{tabularx}{\linewidth}{>{\raggedright\arraybackslash}p{2.3cm}
                             >{\raggedright\arraybackslash}X
                             >{\raggedright\arraybackslash}X
                             >{\raggedright\arraybackslash}X}
\toprule
\textbf{Ethical principle} & \textbf{Governance} & \textbf{Foundation model} & \textbf{Application} \\
\midrule
Beneficence     & Organizational intent and stated mission & Ambiguous due to multipurpose nature & Observable in operational context \\
Non-maleficence & Speculative risks and internal processes & Experimental evaluation and red teaming & Concrete, measurable, and contextualized risks \\
Autonomy        & General policies on responsible interaction & Evaluation of manipulation or induced dependence & Evaluation on real users and concrete effects \\
Justice         & Methodological bias mitigation & Hypothetical analysis of differential behavior & Impact evaluation on concrete groups \\
Explicability   & Organizational documentation and traceability & Limited by model opacity & High auditability through operational traceability \\
\bottomrule
\end{tabularx}
\end{table}

Consequently, the ethical verifiability of AI systems is not homogeneous across layers, and the contemporary conditions of the technology ecosystem make it operationally necessary to shift effective auditing efforts toward the application level, where risk acquires observable, contextualized, and measurable properties, and where responsibilities are subject to organizational, sectoral, and national regulations.

\section{The Proposal}
\label{sec:proposal}

Given the inherent opacity of the governance layer and, partially, of the foundation model itself, this work proposes shifting operational auditing efforts toward those levels where observability and effective control are technically feasible. From this perspective, the foundation model is no longer understood as the final object of evaluation, but as an internal component of a broader sociotechnical system whose emergent behavior depends both on the base model and on the software layers, operational rules, interaction mechanisms, and institutional context in which it is ultimately deployed.

\begin{figure}[p]
\centering
\resizebox{\linewidth}{!}{
\begin{tikzpicture}[
  font=\sffamily\scriptsize,
  >=Stealth,
  txt/.style={align=center},
  hexa/.style={chamfered rectangle, chamfered rectangle xsep=2cm, draw, fill=white,
               align=center, minimum height=0.62cm, inner xsep=2pt, font=\sffamily\tiny},
  db/.style={cylinder, shape border rotate=90, aspect=0.25, draw, fill=white,
             align=center, minimum width=0.95cm, minimum height=0.95cm, inner sep=1pt, font=\sffamily\tiny},
  doc/.style={draw, fill=white, align=center, minimum width=1.25cm, minimum height=1.25cm,
              inner sep=2pt, font=\sffamily\tiny,
              path picture={\fill[white] ($(path picture bounding box.north east)+(-0.25,0.01)$) -- ++(0.26,0) -- ++(0,-0.26) -- cycle;
                            \draw ($(path picture bounding box.north east)+(-0.25,0)$) -- ++(0,-0.25) -- ++(0.25,0);
                            \draw ($(path picture bounding box.north east)+(-0.25,0)$) -- ++(0.25,-0.25);}},
  region/.style={draw, dashed, rounded corners=8pt},
]
\fill[orange!18, draw=black] (0,-0.2) ellipse (6.1cm and 6.4cm);          
\fill[yellow!20, draw=black] (0,2.0) ellipse (3.1cm and 4.1cm);            
\fill[gray!35,  draw=black] (0,4.85) ellipse (2.15cm and 1.0cm);          

\node[fill=blue!75!cyan, text=white, inner sep=3pt] at (0,5.35) {AI provider};
\node[txt] at (0,4.55) {\textbf{$<$ Governance-level audit $>$}\\ Self-regulation\\ Opacity};

\node[txt] at (0,3.2) {\textbf{$<$ Model-level audit $>$}\\ domain-agnostic testing};
\node[txt] at (0,2.25) {CVSS-inspired\\ severity assessment};
\node[db]   (adv) at (-1.95,1.25) {Adversarial\\attacks\\on LLMs};
\node[hexa] (mod) at (0.1,1.25)   {Attack-difficulty\\modulator};
\draw[->, thick] (adv) -- (mod);

\node[txt] at (0.1,0.42) {Prompt wrapper};
\draw[dashed] (-2.65,-0.95) rectangle (2.85,0.18);
\node[hexa] (sc)  at (0.1,-0.12) {Scanning tool};
\node[hexa] (scl) at (-1.75,-0.55) {Scanning tool};
\node[hexa] (scr) at (1.95,-0.55)  {Scanning tool};
\draw[->] (mod) -- (sc);

\node[fill=red!60!black, text=white, minimum width=2.4cm, minimum height=1.35cm,
      rounded corners=3pt, text depth=-0.25cm, txt] (app) at (0.1,-1.65) {Application software};
\node[fill=green!50!black, text=white, minimum width=1.55cm, inner sep=2pt, txt,
      font=\sffamily\tiny] (fm) at (0.1,-1.05) {Foundation\\model};
\draw[->] (sc) -- (fm);
\draw[->] (scl.south) |- ($(app.west)+(0,0.25)$);
\draw[->] (scr.south) |- ($(app.east)+(0,0.25)$);

\node[doc] (r3) at (4.25,2.15)  {Third-party\\risks};
\node[doc] (ru) at (-3.75,-1.25) {Risks to\\users};
\node[doc] (rp) at (3.95,-1.25)  {Risks to the\\IT provider};
\draw[dotted, thick] (scr.east) -- (r3.south west);
\draw[dotted, thick] (scl.west) -- (ru.north east);
\draw[dotted, thick] (scr.east) -- (rp.north west);

\draw[region] (-4.85,-2.75) rectangle (5.25,3.65);
\node[rotate=90, font=\sffamily\small] at (-5.25,0.45) {Black-box evaluation};

\draw[region] (-4.85,-6.55) rectangle (5.25,-3.05);
\node[rotate=90, font=\sffamily\small] at (-5.25,-4.8) {Grey-box evaluation};

\node[db] (ind)  at (-3.1,-3.25) {Industry\\profile};
\node[db] (usr)  at (-2.0,-3.25) {User\\profile};
\node[db] (biz)  at (-2.0,-4.6)  {Business\\knowledge};
\node[hexa, minimum height=0.75cm] (ids) at (0.1,-4.6) {In-domain\\scanning tool};
\draw[->, thick] (biz) -- (ids);
\draw[->, thick] (ids.north) -- (app.south);

\node[doc] (xai) at (2.85,-4.6) {Signed\\XAI reports};
\draw[->, thick] (app.east) -| (xai.north);

\node[txt] at (0,-5.65) {\textbf{$<$ Application-level audit $>$}};
\node[txt, font=\sffamily\bfseries\large] at (0,-6.15) {Known risks};
\end{tikzpicture}}
\caption{Proposed framework.}
\label{fig:framework}
\end{figure}
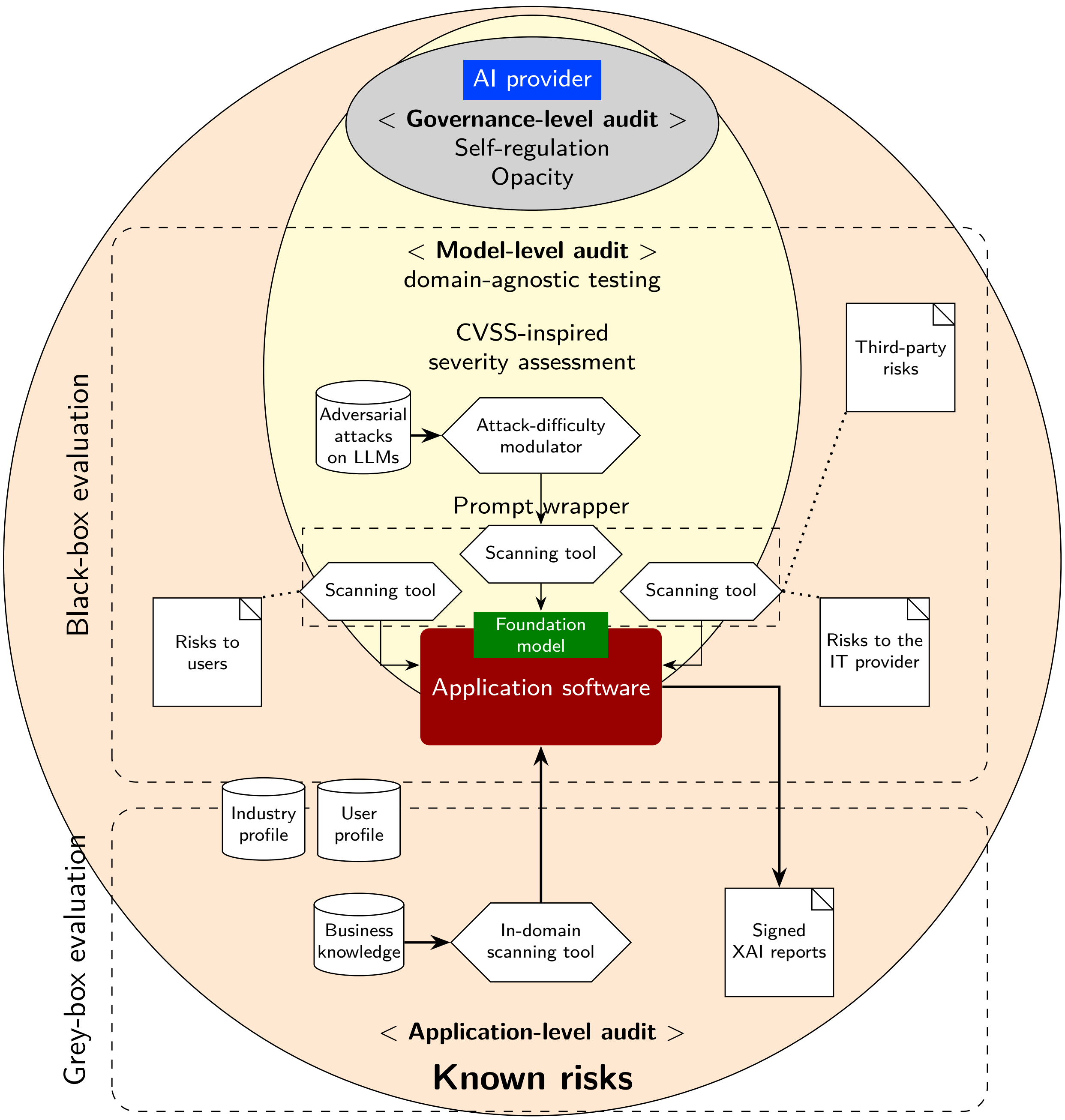

The proposal centers on two complementary forms of evaluation: a black-box evaluation aimed at the general, domain-agnostic analysis of the language model's behavior, and a grey-box evaluation situated on the concrete application and its operational context (Figure~\ref{fig:framework}). Together, both approaches enable a transition from a general perspective on the severity of problems toward a contextualized characterization of operational risk. In this context, it is important to distinguish between severity and risk. The severity of a problem corresponds to general and intrinsic properties of the model's behavior, such as the capacity to manipulate people, generate harmful content, susceptibility to adversarial attacks, or potential psychological impact on users. These properties can be observed through general benchmarking techniques, red teaming, and adversarial exploration in general, even when the specific application domain is still unknown.

However, the existence of severe behavioral problems does not necessarily imply an equivalent level of risk in all deployment contexts. Operational risk emerges when such capabilities interact with real users, institutional processes, sensitive information, and specific application domains. Accordingly, the proposal sets out a structured transition from the general assessment of problem severity toward a contextualized risk assessment based on observable operational impact~\cite{pinacho2025}.

Black-box evaluation is the first level of this strategy and seeks to characterize general behaviors of the foundation model without requiring internal knowledge of its architecture or training. At this level, vulnerabilities, emergent behaviors, harmful content generation, susceptibility to adversarial manipulation, and potentially risky capabilities are explored from a cross-cutting, domain-agnostic perspective. Grey-box evaluation then incorporates partial knowledge of the operational environment of the audited application, including institutional rules, user profiles, business context, known risks, and specific deployment mechanisms. This approach makes it possible to contextualize the observations obtained in the previous phase and transform them into concrete estimates of operational risk.

Under this approach, ethical principles can be evaluated much more specifically: beneficence through explicit knowledge of the system's functional objectives; non-maleficence through contextualized exploration of harm surfaces and operational failures; autonomy through the analysis of patterns of behavioral influence, anthropomorphization, or functional dependence; justice through the evaluation of the concrete impacts of biases on affected groups; and explicability through mechanisms of operational traceability, decision auditing, and generation of verifiable evidence.

The following subsections describe both evaluation levels and their articulation within the proposed framework in detail.

\subsection{Black-Box Evaluation}
\label{sec:blackbox}

Black-box evaluation is the first operational level of the proposed framework and is oriented toward the general characterization of foundation model behavior without requiring internal access to architectures, weights, or development processes. This approach is particularly relevant in the contemporary AIaaS context, where models are usually presented as services whose technical internals are inscrutable.

From this perspective, natural language is understood as a broad interaction protocol capable of exposing the functionalities, restrictions, and emergent behaviors of the system. On this basis, the proposal considers the use of conversational scanning tools available in the ecosystem, such as Garak~\cite{derczynski2024}, Microsoft Counterfit~\cite{counterfit2021}, Promptfoo~\cite{promptfoo2025}, and Giskard~\cite{giskard2025}. These tools make it possible to identify potentially dangerous behaviors through adversarial exploration via red teaming and automated execution of tests aimed at detecting conversational vulnerabilities, harmful content generation, contextual manipulation, and behavioral deviations.

Traditionally, these tools report results based mainly on rates of failed tests or detection of unsafe behaviors. This work proposes complementing that approach with a characterization of the severity of the detected problems inspired by metrics used in software vulnerability assessment, particularly the Common Vulnerability Scoring System (CVSS)~\cite{cvss31}. This approach extends ideas previously explored in~\cite{pinacho2025}, where a CVSS-inspired operational metric was proposed to characterize behavioral vulnerabilities in LLM-based chatbots.

From this perspective, the dangerousness of a behavior depends not only on the potential harm of the generated response, but also on the difficulty required to induce that behavior. Consequently, the appearance of dangerous responses under ordinary conversational interaction, or without explicit manipulation of the system, is considered especially critical, since this type of behavior could inadvertently affect users. Conversely, behaviors that require advanced jailbreaking techniques, prompt injection, or any form of deliberate adversarial manipulation are considered comparatively less severe from a general operational perspective.

To support this evaluation, the proposal incorporates wrappers around the conversational scanning tools, designed to modulate different levels of attempts to induce mission deviation through progressive adversarial techniques. This makes it possible to analyze how the system's behavior varies under different degrees of user-induced manipulation. In addition, and following the approach presented in~\cite{pinacho2025}, the transition from a general characterization of problem severity toward preliminary estimates of operational risk begins even within this black-box stage. To this end, basic contextual factors are incorporated, particularly general user profiles and the sensitivity of the application domain.

In this context, variables such as age, potential susceptibility to misinformation, or exposure to persuasive mechanisms can significantly modulate the expected impact of the same system behavior. Likewise, the industrial sector in which the application operates directly influences the operational criticality of the detected failures, since certain domains, such as health, education, or critical infrastructure, entail potentially more severe consequences in the face of errors, manipulation, or generation of incorrect information.

\subsection{Grey-Box Evaluation}
\label{sec:greybox}

Grey-box evaluation is the second operational level of the proposed framework and is oriented toward making risks visible within the actual deployment environment of the audited application. In this work, the concept of grey-box evaluation does not necessarily imply internal access to the underlying foundation model (whose weights, architecture, or training data may remain completely opaque), but rather the use of contextual information from the client organization, its users, and the industrial domain in which the system operates.

From this perspective, the LLM remains technically a black box. What is actually revealed and evaluated is the sociotechnical system that incorporates the model as an internal component, including its interaction mechanisms, operational rules, orchestration structures, external tools, and associated institutional processes. This approach is especially relevant because the observable behavior of LLM-based applications increasingly depends on additional layers of software and knowledge that go well beyond the isolated foundation model. Systems based on fine-tuning~\cite{hu2021}, Retrieval-Augmented Generation (RAG) architectures~\cite{lewis2020}, conversational memory~\cite{park2023}, prompt orchestration~\cite{chase2023}, external tools, and operational validators significantly modify the generated responses and, therefore, also alter the effective risks of the deployed system.

At this level, the correctness of responses takes on a deeper meaning than the mere surface coherence of the generated content. Tests against misinformation, hallucinations, or inconsistent responses no longer depend exclusively on the parametric knowledge of the model and instead explicitly consider the operational quality of responses within the application domain. This makes it possible to evaluate properties such as contextual coherence, completeness, consistency with institutional sources, and alignment with specific business rules.

To this end, the proposal incorporates in-domain scanning mechanisms that complement the agnostic tests performed during the black-box stage. These evaluations can rely on specific methodologies for RAG systems, such as RAGAS~\cite{es2024} or equivalent approaches, making it possible to inspect not only the final generated response, but also the quality of information retrieval, contextual relevance, and argumentative consistency of the audited system. From this perspective, the transition from general severity toward contextualized operational risk becomes significantly more precise, since the observed behaviors can be evaluated directly on real interaction scenarios, concrete user profiles, and the specific requirements of the organization.

A central element of this proposal is the explicit strengthening of the auditable traceability of the responses generated by the system. In this sense, every relevant interaction should generate, in parallel, a standardized explainability report directly associated with the response produced by the system. This report should contain sufficient evidence to operationally reconstruct the response generation process, including execution context, retrieved sources, components involved, relevant decisions, and potentially critical events.

In addition, these explainability reports should incorporate digital signature mechanisms, providing properties comparable to non-repudiation systems. Under this logic, the goal is not only to provide interpretable explanations for end users, but to build verifiable evidence aimed at subsequent auditing, continuous monitoring, and operational accountability within AI-based sociotechnical systems. Finally, grey-box evaluation makes it possible to extend the traditional paradigm of one-off testing toward continuous oversight schemes. Since LLM-based systems can dynamically change their behavior through configuration changes, model updates, incorporation of new information sources, or changes in operational rules, ethical auditing cannot be understood as a static process. Consequently, continuous operational oversight emerges as an essential component for maintaining verifiability, traceability, and effective control over AI systems deployed in real contexts.

\section{Conclusions}
\label{sec:conclusions}

The contemporary AI landscape is marked by increasing technical opacity, technological concentration, and limited effective regulatory capacity over foundation models and their development processes. Under these conditions, traditional approaches focused exclusively on governance or on internal model inspection are insufficient to guarantee robust ethical and operational oversight of AI systems deployed in real contexts.

This work proposes a pragmatic approach inspired by Zero-Trust principles, shifting effective auditing efforts toward those levels where observability, contextualization, and real control capacity are technically feasible. From this perspective, the foundation model is no longer understood as the sole relevant object of evaluation and becomes part of a broader sociotechnical system whose emergent behavior depends both on the base model and on the operational structures that surround it.

The proposal articulates two complementary levels of evaluation. Black-box evaluation makes it possible to characterize general behaviors, vulnerabilities, and potentially risky capabilities in a domain-agnostic manner, incorporating notions of problem severity inspired by traditional vulnerability assessment approaches. Grey-box evaluation then contextualizes these observations within the actual operational environment of the audited application, enabling a transition from an abstract characterization of severity toward a contextualized assessment of operational risk.

Under this logic, the ethical principles proposed by Floridi and colleagues cease to be merely abstract governance guidelines and become properties that can be operationalized through concrete evaluation, monitoring, and traceability mechanisms. In particular, explicability acquires a cross-cutting role by acting as verifiable support for the practical oversight of beneficence, non-maleficence, autonomy, and justice within real systems. This work also argues for strengthening auditable traceability through mechanisms comparable to non-repudiation systems, proposing digitally signed explainability reports (signed XAI reports) as support for subsequent auditing, continuous monitoring, and operational accountability in LLM-based systems.

Finally, we argue that the ethical auditing of AI systems cannot be understood as a static or exclusively documentary process. The dynamic nature of contemporary foundation-model-based systems demands continuous operational oversight mechanisms capable of adapting to configuration changes, the incorporation of new capabilities, and the constant evolution of the sociotechnical environments in which these technologies are deployed. Thus, rather than replacing future forms of robust governance over foundation models, the proposal seeks to offer a pragmatic operational framework for ethical auditing under the current conditions of opacity, complexity, and rapid technological evolution that characterize today's AI ecosystem.

\section*{Acknowledgments}
This work was partially supported by ANID/FONDECYT Initiation Grant No.~11230359.


\end{document}